\documentclass[prb,10pt,twocolumn,superscriptaddress,notitlepage,floatfix,amssymb]{revtex4}

\usepackage[a4paper,hmargin={1.5cm,1.5cm},vmargin={1.5cm,2.5cm}]{geometry}
\usepackage{graphicx, times}
\usepackage{color}
\usepackage[small,compact,raggedright]{titlesec}

\begin{document}

\newcommand{\ket}[1]{\ensuremath{\left|{#1}\right\rangle}}
\newcommand{\bra}[1]{\ensuremath{\left\langle{#1}\right|}}

\title{Switchable ultrastrong coupling in circuit QED}

\date{\today}

\author{B. Peropadre}
\affiliation{Instituto de F\'{\i}sica Fundamental, CSIC, Serrano 113-bis, 28006 Madrid, Spain}

\author{P. Forn-D{\'\i}az}
\affiliation{Quantum Transport Group, Kavli Institute of Nanoscience, Delft University of Technology, Lorentzweg 1, 2628 CJ Delft, The Netherlands}

\author{E. Solano}
\affiliation{Departamento de Qu\'{\i}mica F\'{\i}sica, Universidad del Pa\'{\i}s Vasco - Euskal Herriko Unibertsitatea, Apdo.\ 644, 48080 Bilbao, Spain}
\affiliation{IKERBASQUE, Basque Foundation for Science, Alameda Urquijo 36, 48011 Bilbao, Spain}

\author{J. J. Garc\'{\i}a-Ripoll}
\email{jjgr@imaff.cfmac.csic.es}
\affiliation{Instituto de F\'{\i}sica Fundamental, CSIC, Serrano 113-bis, 28006 Madrid, Spain}

\maketitle

{\bf Superconducting quantum circuits~\cite{clarke08} possess the ingredients for quantum information processing and for developing on-chip microwave quantum optics~\cite{schoelkopf08}. From the initial manipulation of few-level superconducting systems (qubits)~\cite{bouchiat98,mooij99,martinis85} to their strong coupling to microwave resonators~\cite{blais04,wallraff04,chiorescu04},  the time has come to consider the generation~\cite{houck07} and characterization~\cite{romero09,romero10} of propagating quantum microwaves. In this paper, we design a key ingredient that will prove essential in the general frame: a swtichable coupling between qubit(s) and transmission line(s) that can work in the ultrastrong coupling regime~\cite{abdumalikov08,bourassa09}, where the coupling strength approaches the qubit transition frequency~\cite{ciuti05,guenter09}. We propose several setups where two or more loops of Josephson junctions are directly connected to a closed (cavity) or open transmission line. We demonstrate that the circuit induces a coupling  that can be modulated in strength and type. Given recent studies showing the accesibility to the ultrastrong regime~\cite{abdumalikov08,bourassa09}, we expect our ideas to have an immediate impact in ongoing experiments.}

Quantum circuits is one of the most rapidly developing technologies in quantum information processing. Apart from advances in fabrication and decoherence control, the real boost came through the achievement of the strong coupling regime between qubits and confined microwave photons~\cite{blais04,wallraff04,chiorescu04}. The qubit-cavity couplings reached 10-100 MHz, exceeding a few orders of magnitude the rate at which photons leak the resonator. Recently, with improved techniques and the use of the transmon qubit~\cite{koch07}, the couplings were increased by a factor of 2-3 reaching a strong-coupling regime that can only be compared to the state-of-the-art in of quantum optical microwave domain~\cite{haroche06,brune08}. More recently, proof-of-principle theoretical and experimental studies have allowed the access to the ultrastrong coupling regime~\cite{abdumalikov08,bourassa09}. Here, the coupling approaches the qubit transition frequency and the Jaynes-Cummings model of cavity QED~\cite{haroche06,bourassa09} breaks down~\cite{ciuti05,guenter09}, opening the door to a wide variety of studies around the rather unexplored physics beyond the rotating-wave approximation~\cite{dezela97,hines04}.

The strong coupling regime in circuit QED has made possible an incredible variety of experiments, such as dispersive readouts of qubits~\cite{filipp04}, resolving the photon numbers in cavity~\cite{schuster07}, multiphoton excitations of the Jaynes-Cummings model~\cite{deppe08}, preparing nonclassical states of a resonator~\cite{hofheinz08}, full quantum tomography of the microwave radiation field~\cite{hofheinz09}, or the Tavis-Cummings model in three qubits in circuit QED~\cite{fink09}, etc. However, all those experiments have something in common: the microwave field is confined inside a resonator. The latter induces a discrete spectrum in the transmission lines, such that the coupling between qubits and photons can be effectively switched on and off by tuning the qubit frequency~\cite{mariantoni08} or by using a tunable cavity~\cite{johansson09}. In open transmission lines carrying propagating quantum microwaves, where the qubit couples to a continuous of modes, no method is known to switch off the coupling, be strong or ultrastrong. It is for this reason that open microstrips or transmission lines can be regarded as uncontrolled decoherence sources.

\begin{figure}[t]
\centering
\includegraphics[width=0.76\linewidth]{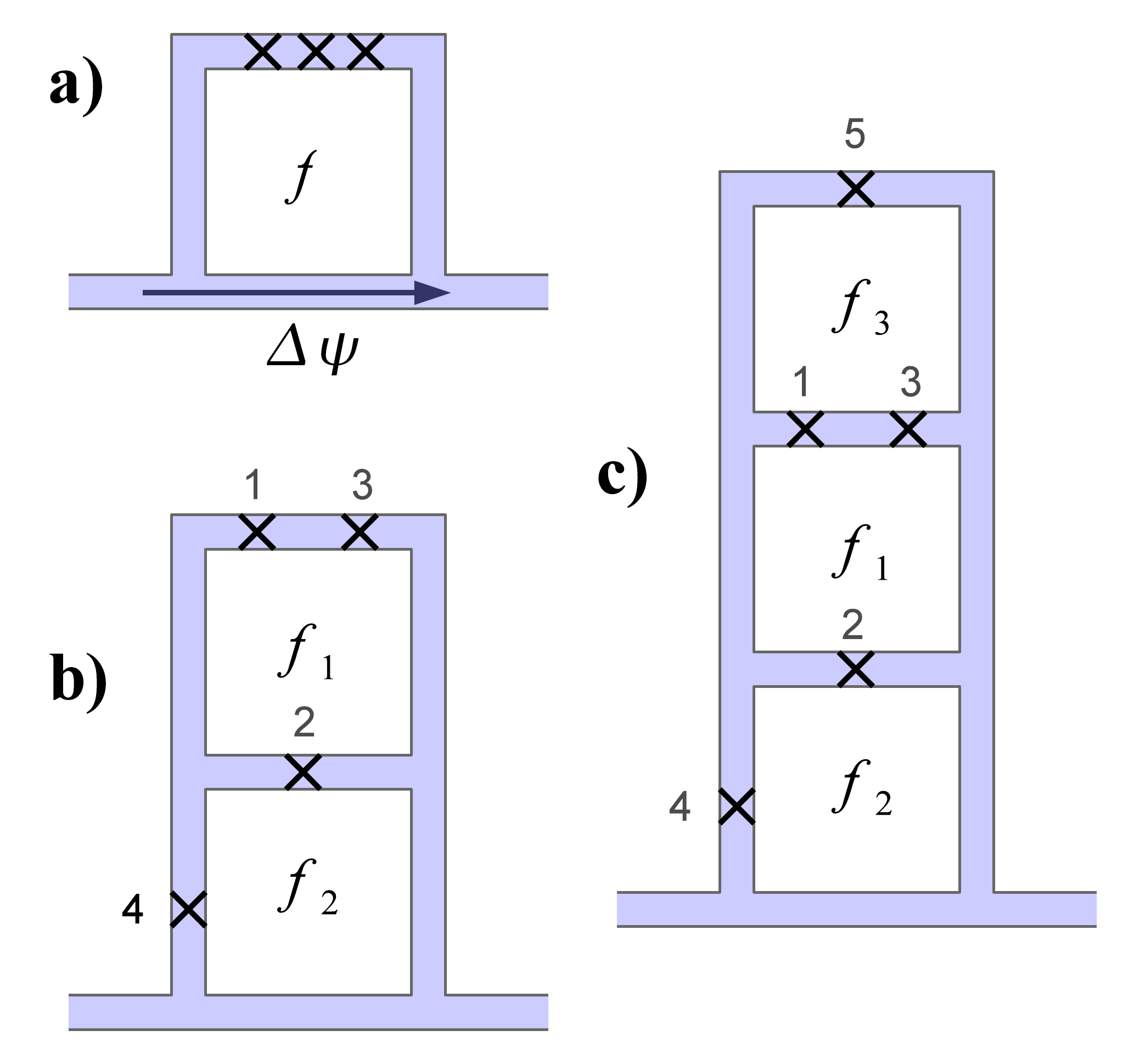}
\caption{Three schemes leading to strong or ultrastrong coupling between a qubit and a transmission line. (a) Basic setup of a qubit coupled directly to the line, $\Delta\psi.$ (b) By adding a second loop, the previous coupling can be modulated. (c) A slightly improved setup in which the qubit is better decoupled from the flux $f_2.$}
  \label{fig:setups}
\end{figure}

In this work, we will introduce a novel circuit QED design where the qubit is ultrastrongly coupled to a transmission line, open or not, with a coupling that can be tuned in strength and kind by applying an external flux bias. Our proposal uses the type of designs shown in Fig.~\ref{fig:setups}, where the qubit is built in direct contact with the transmission line. It has been shown theoretically~\cite{bourassa09}, and demonstrated experimentally\cite{abdumalikov08}, that the system admits an effective description based on a two-level system ---the current in the loop--- ultrastrongly coupled to the photons in the line. We will boost these ideas and show that, by means of induced quantum interference, one is capable of cancelling the ultrastrong coupling, effectively rotating the qubit basis or activating higher-order nonlinearities. This fully controllable coupling tunability opens the path for new experimental results and nontrivial applications. A very important one is switching on and off the interaction in order to resolve the time domain of the qubit evolution with sub-nanosecond resolution, the time needed to apply the switching fluxes. This will allow to resolve the emission and propagation of single photons, measuring their light cone and studying the propagation of entanglement between qubits coupled to the same transmission line~\cite{sabin09}. Straightforward extensions of this work will also allow the implementation of ultrafast quantum switches between cavities and remote qubits, or the design of qutrits with tunable couplings.

\begin{figure*}[t!]
  \centering
  \resizebox{0.93\linewidth}{!}{
  \includegraphics{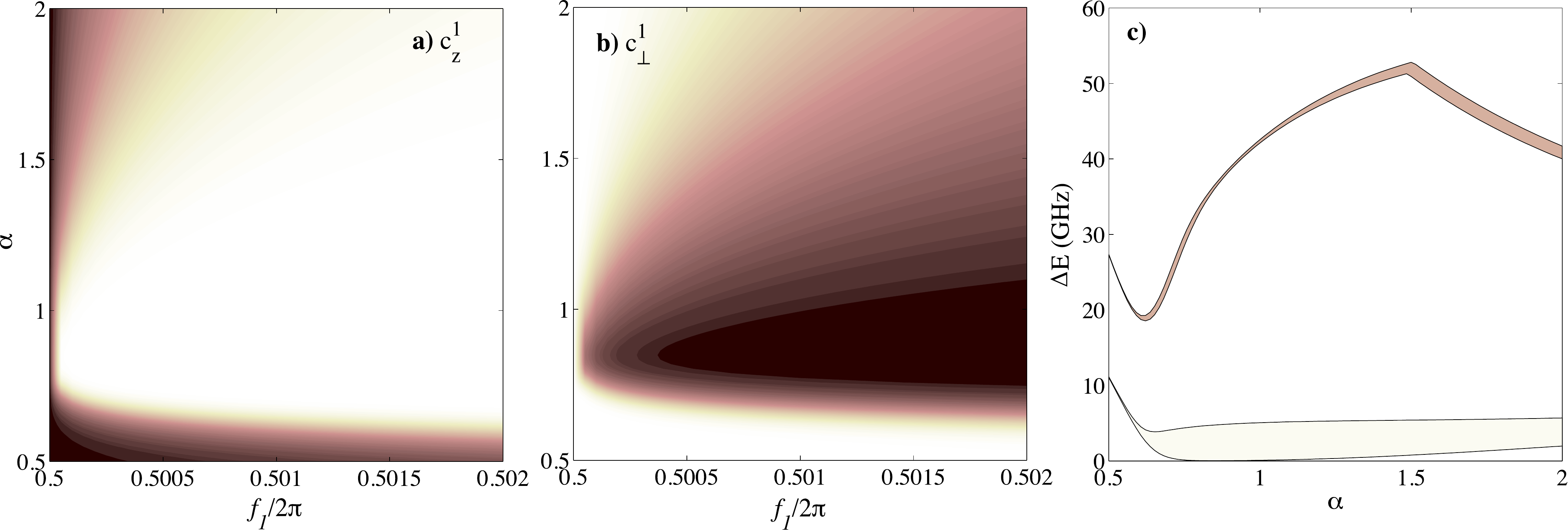}}
\caption{(a-b) For the setup in Fig.~\ref{fig:setups}b, coupling strengths as a function of the external flux $f_1$ and the qubit junction $\alpha,$ for fixed $f_1+f_2=\pi.$ We plot the normalized first order coupling along the Z direction, $c^1_z\sigma_z\Delta\psi,$ and also across the XY plane. (c) Assuming $J=250$ GHz, we plot the ranges of qubit energy differences, $\Omega,$ and the gap to the first excited state, $\Delta E,$ as a function $\alpha$ and for the ranges of fluxes we explored, $f_1 = \pi$ up to $f_1=1.004\pi.$ This plot ensures that we have a qubit for all values of $\alpha$ and $f_1$ considered.}
  \label{fig:setup1}
\end{figure*}

The basic design of the switchable coupling can be understood using a few rules that focus on the inductive terms of the Hamiltonian. More precisely, we will concentrate on the dominant contributions to the energy, which are given by the Josephson junctions as $V(\phi_n)=-J_n\cos(\phi_n).$ Here, $J_n$ denotes the Josephson energy of the $n-$th junction and $\phi_n$ is the phase difference between both sides of the junction. These phases are by the Josephson relation proportional to the flux across the device, $\phi=\varphi/\varphi_0$ with the unit of flux $\varphi_0=\hbar/2e.$ Next, around close loops the total flux is quantized and a multiple of $h/2e.$ This quantization imposes relations between the flux jumps on different junctions, reducing the complexity of the problem, $\sum_n\phi_n = f + 2\pi\times n$ but it also introduces a control parameter which is the externally applied magnetic flux inside the loop, $f\varphi_0.$ Finally, we will include an additional flux difference, $\Delta\psi=\Delta\varphi/\varphi_0,$ along the segment that is shared with the transmission line (see Fig.~\ref{fig:setups}a) and which is the source of the coupling.

With these rules, one can analyze the setup from Fig.~\ref{fig:setups}a and impose the usual flux qubit configuration, with two equal junctions $J_1=J_3=J,$ and a smaller one $J_2=\alpha J~(\alpha<1),$ and the quantization $\phi_1 + \phi_2 + \phi_3 - \Delta\psi = f + 2\pi\times n$. The result is an effective Hamiltonian that, for $f=\pi$, reads
\begin{eqnarray}
  H_{J}&=&-J\cos(\phi_1) - \alpha J\cos(\phi_2) - J\cos(\phi_3) \\
  &=& J\left[\alpha \cos(\phi_+) - 2\cos(\phi_-/2)\cos(\phi_+/2)\right]  \nonumber \\
  && + \, \alpha J \Delta\psi \sin(\phi_+) + {\cal O}(\Delta\psi^2) \, . \nonumber
\end{eqnarray}
Note how this model combines a flux qubit term~\cite{mooij99}, where the most important variable is the linear combination $\phi_+=\phi_3+\phi_1,$ with a coupling between the qubit degrees of freedom and the transmission line. When we introduce the capacitive terms the qubit can be diagonalized and the previous model becomes
\begin{eqnarray}
  H &\sim& {\textstyle\frac{1}{2}}\Omega \sigma_z + \alpha J \Delta\psi \sigma_x \, .
\end{eqnarray}
It is noteworthy to mention that the qubit-line coupling can remain in the ultrastrong regime\cite{bourassa09}, because it is proportional to the Josephson energy, $\alpha J.$ However, the coupling mantains the form $\sigma_x \Delta \psi$ and there is not enough tunability because the junction that determines the coupling, $J_2,$ is also responsible for the existence of the qubit.

A more versatile design, shown in Fig.~\ref{fig:setups}b, separates the three qubit junctions and the transmission line by a loop. The new Josephson junction adds a contribution to the energy which is of the form $J_4\cos(\phi_4)=\alpha_4J\cos(f_2-\phi_2-\Delta\psi),$ while keeping the flux qubit quantization independent of the transmission line flux, $\Delta\psi.$ The result is now
\begin{eqnarray}
  H&=& J\left[-\alpha \cos(f_1-\phi_+) - 2\cos(\phi_-/2)\cos(\phi_+/2)\right] + \label{model2} \nonumber\\
  && +\alpha_4J \cos(f_1+f_2-\Delta\psi+\phi_+) \, ,
\end{eqnarray}
which contains two independently adjustable parameters, $f_1$ and $f_1+f_2.$ If we adjust $f_2=\pi/2$ to stay within the linear coupling with $\Delta\psi,$ the Hamiltonian reads
\begin{equation}
  H \sim {\textstyle\frac{1}{2}}\Omega \sigma_z +
  \alpha_4J \Delta\psi \sum_{r=x,y,z} c^1_r \sigma_r \, ,
\end{equation}
and has a tunable orientation in the qubit basis.

Moreover, since the coupling term is strictly independent of the qubit Hamiltonian, it now becomes possible to switch on and off of the interaction. The simplest way is to replace the fourth junction, $J_4,$  with a SQUID, so that a control flux over this loop will allow us to dynamically tune the coupling strength, $\alpha_4.$ Using this technique the mutual influence between the qubit and the transmission line can be completely suppressed in times of about 0.1 ns, that is much faster than the qubit-resonator dynamics. Remark that in the ultrastrong coupling regime the rotating-wave-approximation cannot be made, and the physics of Rabi oscillations does not apply.

\begin{figure*}[t!]
  \centering
  \resizebox{0.93\linewidth}{!}{
  \includegraphics{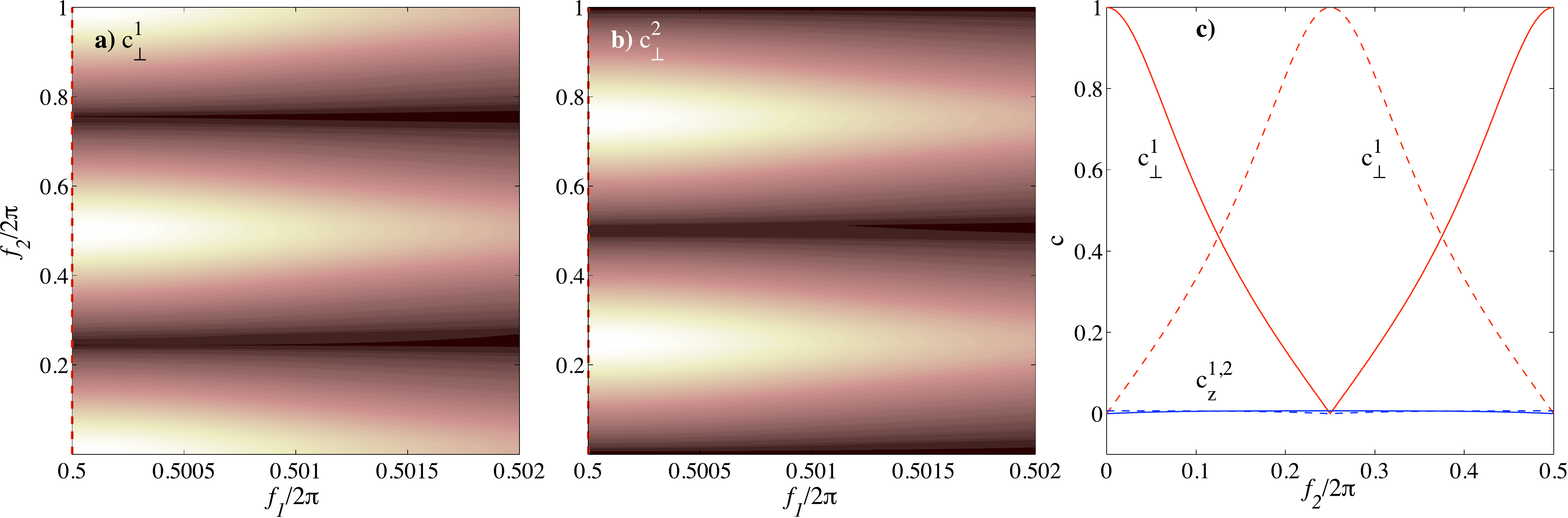}}
 \caption{(a-b) Normalized transverse couplings, $c^{1,2}_x,$ as a function of external fluxes $f_1$ and $f_2$ for the setup in Fig.~\ref{fig:setups}c, using $\alpha=2.0$ and $\alpha_4=0.1.$ (c) Cut at $f_1=0.5$ shows first-order (solid) and second-order couplings (dashed) of longitudinal ($c_z,$ blue) and transverse ($c_x$ red).}
  \label{fig:setup2}
\end{figure*}

A different setup which we consider in this work is shown in Fig.~\ref{fig:setups}c. We now included two equal junctions, $J_5=J_4=\alpha_4J,$ which are fixed and the previous control flux for the junction was replaced by an additional flux across the first loop $f_3=\pi-f_2-f_3.$ The value of the flux is designed to cancel the zero-th order term which appears in Eq.~(\ref{model2}) when we move away from $f_2=\pi/2,$ and together with $f_1$ and $f_2$ it permits (i) switching on and off the interaction, (ii) changing the orientation and (iii) increasing the relevance of higher order couplings. The effective Hamiltonian now reads
\begin{equation}
  H = \frac{1}{2}\Omega\sigma_z + \alpha_4 J \sum_n \Delta\psi^n \sum_{r=x,y,z} c^{n}_r\sigma_r.
\end{equation}

We have analyzed these setups numerically, confirming that the coupling is ultrastrong and can be arbitrarily tuned. In order to do so, we first completed the theoretical model to include the capacitive terms which appear in the junctions and the line itself. We then diagonalized the Hamiltonian of what we identify as the qubit degrees of freedom, obtaining the two lowest energy states and verifying that these states are protected from higher energy excitations [Fig.~\ref{fig:setup1}c]. Finally, we expanded the interaction between the qubit and the transmission line in powers of the flux $\Delta\psi,$ and computed the matrix elements of the interaction in the qubit basis.

The main results are shown in Fig.~\ref{fig:setup1} and Fig.~\ref{fig:setup2}, which correspond the two respective setups in Figs.~\ref{fig:setups}b and Figs.~\ref{fig:setups}c. In the first figure we have explored the simplest switchable setup for various configurations of the qubit, $\alpha,$ and of the externally applied flux, $f_1.$ It is important to see that we have a very good qubit for values of $\alpha$ well above the $0.8$ which is normally considered. Furthermore, when $f_1=\pi$ both for $\alpha<1$ and $\alpha>1$ the ground states are superpositions of left- and right-moving currents, and the interaction is proportional to $\sigma_x,$ transversely to the qubit basis. When we apply a small flux difference, increasing or decreasing $f_1$ we unbalance the populations of the two current states, the ground state acquires an effective magnetic dipole and the interaction rotates from $\sigma_z$ to $\sigma_{x,y}.$

The second set of plots is shown in Fig.~\ref{fig:setup2} and corresponds to the three loops setup, Fig.~\ref{fig:setups}c. We have chosen $\alpha=2$ because it allows for a finer control in the rotation of the interaction $\sigma_x$ to $\sigma_z,$ but it is not essential. The tunability of the qubit manifest as follows: when $f_1$ is increased the strength of $c^{1,2}_x$ decreases, causing an increase of $c^{1,2}_z,$ much like in Figs.~\ref{fig:setup1}a-b. But in addition to this, we now have complete freedom to change the value of $f_2.$ Changes in this second flux result in a simultaneous deactivation of all couplings, $c_{x,y,z},$ which become zero as seen in the dark horizontal stripes for $f_2=(2n+1)\pi/4$ in Figs.~\ref{fig:setup2}a-b, and in the zeros of $c_x$ in Fig.~\ref{fig:setup1}c. The switching capability, measured as $\min c_x/c_z,$ is rather strong, $6\times 10^{-4}$ in this example, improves by increasing $\alpha.$

So far we have demonstrated the posibility of achieving a highly tunable qubit-line system. The only question that remains is the strength of that coupling. For clarity, we will restrict to the case in which the line forms a single-mode resonator. The flux gradient then becomes approximately\cite{bourassa09}
\begin{eqnarray}
  \Delta\psi &=& \frac{ \partial u(x) \Delta x}{\varphi_0}
  \sqrt{\frac{\hbar}{\omega C}} (a+a^\dagger)
  = \frac{2\pi\partial_x \psi(x)}{\Phi_0} (a+a^\dag).\nonumber
\end{eqnarray}
Here, $u(x)$ is the wavefunction of the photons inside the cavity, $\Delta x$ is the separation between the two qubit-line intersections, $\omega$ is the cavity mode frequency and $C$ the resonantor total capacitance. The dependence is thus similar to previous works meaning that we can achieve comparable ultrastrong couplings. Assuming a flux gradient\cite{bourassa09} $|\partial_x\psi|=65\times 10^{-6}\Phi_0/\mu\mathrm{m},$ and a qubit size $\Delta x=5\mu\mathrm{m},$ we reach a coupling $g=2 \times 10^{-3} \times J,$ which for a typical junction with $J=250$ GHz implies a very strong $500$ MHz coupling. The previous numbers are however pessimistic. A typical Aluminium penetration depth of 150nm allows a larger flux gradient, of $1.7\times 10^{-3} \Phi_0/\mu\mathrm{m}$ or 25 times the previous coupling strength, that is up to 10 GHz. Either with these values, or enhancing the phase gradient with the use of an auxiliary junction\cite{bourassa09}, the fact is one can take the coupling strength deep in the ultrastrong regime with an interesting consequence, namely the possibility of inducing non-linearities in the transmission line~[Fig.~\ref{fig:setup2}]. In the crudest approximation, the second order coupling strength is proportional to $\alpha_4J(2\pi\Delta x\partial_x\psi/\Phi_0)^2.$ For a phase slip of 0.01-0.03, that means a coupling $J\times( 10^{-4}-10^{-3}),$ or 25 to 250 MHz, according to the values mentioned before.

We have not considered in this work the influence of capacitive couplings. They will typically take the form
\begin{equation}
  H_{\mathrm{cap}} = \frac{\alpha_4}{1+2\alpha+4\alpha_4}2\pi\hbar\omega \frac{\partial_x\psi}{\Phi_0} \Delta x
  \times i(a - a^\dag) Q,
\end{equation}
where $Q=(-i\partial/\partial\phi_+)$ is the charge operator on the coupling junction. A similar analysis gives now a value proportional to $1.8\times 10^{-3} \hbar\omega,$ yielding negligible couplings of at most 1~MHz.

Summing up, in this work we have suggested the possibility of engineering an ultrastrong switchable coupling between a flux qubit and a transmission line. The design, tunability and strength of the coupling does not depend on detuning the qubit, but rather on a suitable coupling and one or more tuning fluxes. The coupling intensity is large enough to envision different possible applications, some of which we summarize now.

The first and most immediate application would be to perform quantum gates between arbitrary pairs of qubits of a row coupled to a transmission line. By decoupling all qubits except those chosen to peform a two-qubit gate, it should be possible to perform operations as the swap of quantum information between the qubit and the line modes, or between both qubits. This scheme has an important advantage, namely that the qubit switching happens for precise values of the fluxes, depending only on geometric properties and not on the precise eigenenergies or fabrication properties of the junctions. A second application would be decoupling a qubit from its environment (the transmission line) and coupling it with slower measurement devices, especially after having performed an ultrastrong coupling evolution~\cite{zeno09}. Furthermore, since the coupling may be switched on and off in about 0.1 ns, this enhanced resolution can also be used for the measurement of quantum microwaves as well. More precisely, given that one qubit may act as a perfect mirror for individual photons, a combination of one or more may be used as streak camera for stroboscopic measurements of wavepackets. A fourth application is the deterministic generation of propagating single- and two-photon pulses. This would work by decoupling the qubit, exciting it, and then activating an ultrastrong coupling dynamics. The qubit would decay in a few nanoseconds, either emitting a single photon (linear coupling) or two of them (nonlinear one). Furthermore, the shape of the photon wavepacket can be tailored using a second qubit that acts as a switchable mirror, as explained before.

In conclusion, we believe that a future access to the physics of switchable ultrastrong coupling will pave the way to novel and otherwise inaccessible physics, without discarding key applications to quantum microwave technologies in circuit QED.

P.F.-D. acknowledges support from NanoNed. E.S. acknowledges funding from UPV-EHU Grant GIU07/40 and EuroSQIP European project. J.J.G.-R. and B.P. thank funding from Spanish MEC Project FIS2006-04885 and CSIC Project 200850I044, CSIC JAE-PREDOC2009 Grant.

\end{document}